 \documentstyle[12pt]{article}
 \newcounter{multieqs}

\begin{document}
 \begin{titlepage}
 \hfill {\Large DTP-MSU 95--01, January 1995}
 \vspace{3cm}
 \begin{center}
 {\LARGE \bf  Stringy Sphalerons and Gauss--Bonnet Term}\\
 \vspace{8mm}
 {\bf Evgeni E. Donets,}\\
 {\it Laboratory of High Energies,\\
 Joint Institute for Nuclear Research, 141980 Dubna, Russia,\\
 e--mail: donets@lhe26.jinr.dubna.su }\\
 and\\
 {\bf Dmitri V. Gal'tsov,}\\
 {\it  Department of Theoretical Physics, Physics Faculty, \\
    Moscow State University, 119899 Moscow, Russia,\\
  e--mail: galtsov@grg.phys.msu.su }\\
 \end{center}
 \vspace{15 mm}
 \begin{center}
 {\bf Abstract}
 \end{center}
 \vspace{5mm}

  The effect of the Gauss--Bonnet term on the $SU(2)$
 non--Abelian regular stringy sphaleron solutions is studied
 within the non--perturbative treatment. It is found that the existence of
 regular solutions depends crucially on the value  of the numerical factor
 $\beta$ in front of the Gauss--Bonnet term in
 the four--dimensional effective action.
 Numerical solutions are constructed in the $N=1, 2, 3$ cases for different
 $\beta$ below certain critical values  $\beta_N$ which decrease with
 growing $N$ ($N$ being the number of nodes of the Yang--Mills function).
 It is proved that  for any static spherically symmetric asymptotically
 flat regular solution the ADM mass is exactly equal to the dilaton charge.
 No solutions were found for  $\beta$  above critical values,
 in particular, for  $\beta=1$.

 \end{titlepage}

 \newpage

   Since the Bartnik and McKinnon's discovery \cite{bm} of the regular
 particle--like solutions to the coupled system of the Einstein--Yang--Mills
 (EYM) equations there was a growing interest in revealing their possible
 physical significance. It was shown that these solution could play at
 the ultramicroscopic distances a role analogous to that of electroweak
 sphalerons \cite{gv}. Sphaleron interpretation is supported by the existence
 of the odd--parity YM negative modes \cite{onm}
 (apart from the previously known
 even--parity ones \cite{sz}), as well as the fermion zero modes and the
 level--crossing phenomenon \cite{fer}. Natural question arises whether the
 EYM sphalerons survive in more sophysticated
 field models suggested by the theory
 of superstrings. It was shown recently that regular sphaleron
 solutions exist within the context of the Einstein--Yang--Mills--Dilaton
 (EYMD) theory \cite{dg1}, \cite{lm}, \cite{tm}, \cite{dg2}, \cite{oneill}.
 Remarkably, they have a dilaton charge exactly
 equal to the ADM mass. This property is similar to that of the extremal
 dilaton black holes which are likely (at least some)
 to represent exact solutions of the string theory.

 To further investigate possible relevance of the EYMD sphalerons to the
 string theory we study here the EYMD system with the Gauss--Bonnet (GB) term
 which is typically present in stringy gravity as the lowest order
 curvature correction. Similar problem for Abelian dilatonic black holes
 was studied recently within the perturbative approach \cite{msm}.
 However, for regular solutions, one needs a more precise treatment.
 In order to see in a continuous way how EYMD solutions are
 modified by the GB term we introduce into the lagrangian a numerical
 factor $\beta$ so that $\beta=0$ corresponds to the pure EYMD system.
 It turns out that series expansion of the regular
 solution near the origin is essentially $\beta$--dependent.
 Also, computing the GB contribution into the effective
 energy density on the background EYMD solutions, one can observe that
 the GB effect becomes non--small for $\beta$ of the order of unity.
 For this reason we avoid any perturbative treatment of the GB term
 and attack the problem numerically. Starting with $\beta=0$ we increase
 gradually the value of this parameter and search
 (using the shooting strategy) for solutions interpolating smoothly
 between the regular asymptotic expansion near the origin and an
 asymptotically flat expansion at infinity. Although the leading terms
 of expansions near infinity are not modified by the GB corrections,
 those near the origin {\em are affected} substantially. We construct
 numerical solutions for $N=1, 2, 3$ and some $\beta \neq 0$ and show
 that regular solution cease to exist above certain critical values
 $\beta_N$ depending of the number of nodes $N$ of the YM function.
 For all solutions found within the domains of existence, modifications
 due to GB term are relatively small, and all characteristic functions
 still preserve the typical behaviour they have in the pure EYMD case.
 We also prove analytically that the dilaton charge of any regular solution
 (with an exact account for the GB term)  {\em is equal} to its ADM mass
 independently on the value of $\beta$. For $\beta=0$ a stronger relation
 holds between $g_{00}$ and the dilaton factor everywhere.

  We start with the following bosonic part of the heterotic string
 effective action in four dimensions in the Einstein frame  :
 \begin{equation}
 S = \frac{1}{16\pi} \, \int \;\left\{(-{\it R} + \, 2\partial_{\mu} \Phi
 \partial^{\mu} \Phi ) - \alpha'\exp (-2 \Phi) (F_{a\mu\nu} \, F_a^{\mu\nu}
 - \beta {G})\right\}\, \sqrt{-g} d^4x \; ,
 \end{equation} 
 \noindent
 where $\Phi$ is the dilaton, $F$ is the Yang-Mills field strength and
 $ G$ is the Gauss--Bonnet term which can be presented as the divergence
 of the topological current
 \begin{equation}
  G = R_{\mu\nu\lambda\tau} R^{\mu\nu\lambda\tau} -
  4 R_{\mu\nu} R^{\mu\nu} + R^2 = \nabla_{\mu}
  K^{\mu} \; .
 \end{equation} 

 Integrating by parts the GB term in (1) one can rewrite the action in
 somewhat simpler form (both in (1) and (2) we ignore surface terms
 which are not relevant for the present analysis)
 \begin{equation}
 S = \frac{1}{16\pi} \, \int \, \,\left\{((-{\it R} + \, 2\partial_{\mu}
 \Phi
 \partial^{\mu} \Phi ) - \alpha' e^{-2 \Phi} (F_{a\mu\nu} \, F_a^{\mu\nu}
 - 2 \beta (\partial_{\mu} \Phi) { K^{\mu}})\right\}\,\sqrt{-g} d^4x \; .
 \end{equation} 
 We parametrize the metric of the static spherically symmetric spacetime as
 \begin{equation}
 ds^2 = W dt^2 - \frac{dr^2}{w} - R^2 (d\theta^2 + \sin^2 \theta d\phi^2)\; ,
 \end{equation} 
 where  $ W = w \sigma^2$ and all functions depend on the single variable
 $r$. In this case only the radial component of the
 topological current is relevant
 \begin{equation}
  K^r = \frac{4 (w \sigma^2)' (wR'^2 - 1)}{R^2\sigma^2} \;.
 \end{equation} 
 Here and below primes mean derivatives with respect to $r$.

 A magnetic part of the static spherically symmetric $SU(2)$ Yang--Mills
 connection can be expressed in terms of the single function of the radial
 variable $f(r)$
 \begin{equation}
 A^a_{\mu} T_a dx^\mu = (f-1)(L_\phi d\theta - L_\theta\sin \theta d\phi)\, ,
 \end{equation} 
 \noindent
 where  $L_r =T_a n^a$,  $L_\theta=\partial_\theta L_r$,
 $L_\phi= (\sin \theta)^{-1}\partial_\phi L_r\;,$
 $n^a = (\sin \theta \cos \phi, \sin \theta \sin \phi, \cos \theta)$
 is the unit vector and $T_a$ are normalized Hermitean generators of
 the $SU(2)$ group.

 Integrating out the angular variables in (2) and eliminating some total
 derivatives one obtains the following reduced effective action
 \begin{equation}
  S = \int dt dr \left(L_{g} + L_{m} + L_{ GB} \right)\; ,
 \end{equation} 
  where
 \begin{equation}
 L_{g} = \frac{\sigma}{2} \left( R'(wR)' + 1\right) +
 w \sigma'RR'\; ,
 \end{equation} 
  is the gravitational part,
 \begin{equation}
 L_{m} = - \frac{1}{2} wR^2 \Phi'^2 - \frac{\alpha'}{2} { F} e^{-2\Phi}\; ,
 \end{equation}   
 is the matter part,
 \begin{equation}
 L_{GB} = 2 \alpha' \beta \sigma^{-1} \Phi' W' (wR'^2 - 1)
 e^{-2\Phi}\; ,
 \end{equation} 
 is the Gauss--Bonnet contribution, and
 \begin{equation}
 { F} = 2wf'^2 + \frac{(1 - f^2)^2}{R^2}\; .
 \end{equation} 

  Note that an arbitrary rescaling of the slope
 parameter $\alpha' \rightarrow k \alpha'$
 together with the corresponding rescaling of the radial variable
 $r \rightarrow \sqrt{k} r$ is a symmetry transformation of the
 effective action (7). Choosing Planck units $\alpha'=1$
 we are left with the only dimensionless parameter $\beta$.

 The equations of motion (including an Einstein constraint) can
 be obtained by direct variation of (7) over $\sigma, w, R, f, \Phi$.
 Then fixing the gravitational gauge as $R=r$ one finds the following
 set of equations

 \begin{equation}
 \frac{\sigma'}{\sigma} = r\Phi'^2 + \frac{2f'^2 e^{-2\Phi}}{r} +
 \frac{4\beta}{r}\Big((\frac{\Phi'(w-1) e^{-2\Phi}}{\sigma})' \sigma -
 \frac{W' \Phi'}{\sigma^2} e^{-2\Phi} \Big)\; ,
 \end{equation}

 \begin{equation}
 w'\left(1 - \frac{4 \beta \Phi' (1 - 3w) e^{-2\Phi}}{r}\right) +
 \frac{{ F}}{r}e^{-2\Phi} + rw \Phi'^2 = \frac{(1 - w)}{r}
 \left(1 - 4 \beta w (e^{-2\Phi})''\right)\; ,
\end{equation}

\begin{equation}
\frac{1}{2}\left(\frac{W'}{\sigma}\right)' r +
\left(\frac{W}{\sigma}\right)' +
\sigma \left(wr\Phi'^2 -\frac{(1 - f^2)^2}{r^3} e^{-2\Phi}\right) +
4 \beta \left(\frac{W'w \Phi'e^{-2\Phi}}{\sigma}\right)' = 0 \; ,
\end{equation}

\begin{equation}
 \left(w \sigma f' e^{-2\Phi}\right)' + \frac{\sigma f (1 - f^2)
e^{-2\Phi}}{r^2} = 0\; ,
\end{equation}

\begin{equation}
\left(\sigma r^2 w \Phi'\right)' + \sigma { F} e^{-2\Phi} +
2 \beta \left(\frac{W'(1 - w)}{\sigma}\right)' e^{-2\Phi} = 0\; .
\end{equation}

   It is useful also to compute an effective energy density as it enters
the standard Einstein equations with account for the GB term
\begin{equation}
2 T^0_0 = w \Phi'^2 + \frac{{ F}}{r^2} e^{-2\Phi} +
\frac{4 \beta}{r^2}\left\{2w(w -1)(\Phi'' - 2\Phi'^2) +
\Phi' w'(3w - 1)\right\}
e^{-2\Phi}\; .
\end{equation}
For $\beta=0$ the system reduces to that of \cite{dg1} and
the corresponding solutions exibit typical BK structure of the YM function:
solutions start from $f=\pm 1$ and goes asymptotically to $\mp 1$ either
monotonically $(N=1)$ or after $N-1$ oscillations around zero.

As a first step of the analysis we calculate the GB term and
the corresponding density $\sqrt{-g}G$ substituting the sphaleron solutions
found without an account for the GB term. Numerical results are shown on the
Fig. 1. One can see that the value of GB term increases with growing
number of nodes of the YM function. It can be anticipated that its
influence on the sphaleron solutions will increase for higher $N$.
We have also calculated the effective energy density (17) for the background
EYMD solutions. Fig. 2 clearly shows that relative contribution of
the GB term for $\beta=1$ is not small. This presumably invalidate
any attempt to treat the GB term perturbatively, so we are faced with
the problem of constructing numerical solutions to the system (12)--(16).

To define the ADM mass $M$ and the dilaton charge $D$ one writes asymptotic
expansions for $W$
\begin{equation}
W = 1 - \frac{2M}{r} - \frac{2D^2 M}{r^3} + O(\frac{1}{r^4})\; ,
\end{equation}
and the dilaton
\begin{equation}
\Phi = \Phi_{\infty} + \frac{D}{r} + \frac{DM}{r^2} +
\frac{8M^2 D - D^3}{6r^3} + O(\frac{1}{r^4})\; .
\end{equation}
The corresponding expansion of $\sigma$ reads
\begin{equation}
\sigma = 1 - \frac{D^2}{2r^2} - \frac{4D^2 M}{3r^3} + O(\frac{1}{r^4})\; .
\end{equation}

To ensure asymptotic flatness it is sufficient (as it is for $\beta=0$)
to have for the  Yang--Mills function $f$
\begin{equation}
f = \pm 1 + O(\frac{1}{r})\; .
\end{equation}
Clearly,  GB--induced terms do not influence the leading behaviour
of solutions near infinity.

In contrary, an expansion of regular solutions near the origin {\em is}
affected by the curvature terms. From the system (12)--(16) one finds
 \begin{eqnarray}
 f&=& -1 + br^2 + O(r^4)\,,\\
 \Phi&=& \Phi_0 + \Phi_2 r^2 + O(r^4)\,,\\
 \sigma&=&\sigma_0 + \sigma_2 r^2 + O(r^4)\,,\\
 W&=&W_0 + W_2 r^2 + O(r^4)\,,
 \end{eqnarray} 
 \noindent
 or in terms of $w$:
 \begin{equation}
 w = 1 + w_2 r^2 + O(r^4)\,,
 \end{equation} 
 where the following relations hold
 \begin{equation}
 W_0 = \sigma_0^2, \quad
 W_2 = 2 \sigma_0 \sigma_2 + w_2 \sigma_0^2 .
 \end{equation} 

 Let us prove that for any regular solution to the system (12)--(16)
 (if exists), the ADM mass $M$ is exactly equal to the dilaton charge $D$.
 Combining Eqs. (12), (14) and (16), after some rearrangment
 one can find the following identity
\begin{equation}
\left(2 \sigma r^2 w \Phi' + \frac{W' r^2}{\sigma}\right)' = 4 \beta Q'\;,
\end{equation} 
where
\begin{equation}
Q = \sigma^{-1} \left\{(w - 1)(W' + 2W \Phi') - 2r \Phi w W\right\}\; .
\end{equation} 
Integrating this relation over the  semiaxis with account for (18)--(20)
on gets
\begin{equation}
\left.\left(\frac{W}{\sigma}r^2 \Phi' +
\frac{W' r^2}{2\sigma}\right)\right|^{\infty}_{0}\equiv M - D =
2 \beta \left[Q(\infty) - Q(0)\right]\; .
\end {equation} 
Now from the expansions (18)--(21) and (22)--(26) it can be found that both
above boundary values of $Q$ are equal to zero, what proves the
exact equality $M=D$. Remarkably, this property of regular EYMD
solutions observed first in \cite{dg1}, remains true with account
for the GB term for any value of $\beta$.
There is an important difference, however. In the case $\beta=0$
a stronger identity
\begin{equation}
W = \exp (-2\Phi)
\end{equation}
holds, which  is similar to the well--known relation for the extremal
magnetic dilatonic black holes, where it
ensures regularity of the metric in the string frame.
When GB term is taken into account this is no longer true while the
relation $M=D$ exibiting the validity of (31) in the asymptotic region
still holds.

  Similarly to the system of Einstein--Yang--Mills--Dilaton equations
  \cite{dg1}, \cite{lm}, \cite{tm}, \cite{oneill}
 without Gauss--Bonnet term there are three independent
 parameters in the series solutions of the system (12--16) near the origin:
 $b$, $\Phi_0$ and $\sigma_0$.
 From them the quantity $\Phi_0$ is somewhat trivial
 because of the symmetry of the system under a dilaton shift accompanied
 by suitable rescaling of the radial coordinate (if desired,
 $\exp (-2\Phi_0)$ may be absorbed into redefinition of parameters in
 (22)--(26)).
 However, there is a substantial complication as compared with the
 pure EYMD theory. In order to fulfil the system (12--16) in the first
 leading order, the coefficient
 $\Phi_2$ has to be one of the real roots of the following algebraic
 equation of the forth order
 \begin{equation}
 \left(\Phi_2 + 2 b^2 e^{-2\Phi_0}\right)
 \left(1 + 16 \beta \Phi_2 e^{-2\Phi_0}\right)^3 +
 32 \beta b^4 e^{-6\Phi_0} \left(1 + 8 \beta \Phi_2 e^{-2\Phi_0}\right) = 0 .
 \end{equation} 
 \noindent
 Once $\Phi_2$ is found, two other coefficients $w_2$ and $\sigma_2$ can
 be obtained as
 \begin{eqnarray}
 w_2 &=& - \frac{4b^2 e^{-2\Phi_0}}{1 + 16 \beta \Phi_2 e^{-2\Phi_0}}\,,\\
 \sigma_2 &=& \frac{\sigma_0 e^{-2\Phi_0}
 (4b^2 + 4 \beta w_2 \Phi_2)}{1 + 16 \beta \Phi_2 e^{-2\Phi_0}}\;  .
 \end{eqnarray} 

 It is convenient to regard the Eq. (32) as giving the value of $\Phi_2$
 as a function of $b$, while $\Phi_0$ is fixed. In fact, a dilaton shift
 \begin{equation}
 \Phi_0 \rightarrow \Phi_0 + \delta \Phi_0
 \end{equation}
 leads to a solution related with the initial one by a radial rescaling.
 Physically the normalization $\Phi_{\infty}=0$ is preferable since it
 ensures a unique mass scale for all solutions. But technically is is
 convenient to solve the system first by fixing $\Phi_0$ arbitrarily,
 say, $\Phi_0=0$. Then the rescaled solution will result from
  \begin{equation}
b \rightarrow b \exp (2\delta \Phi_0) ,\quad
\Phi_2 \rightarrow \Phi_2 \exp (2\delta \Phi_0) ,\quad
\sigma_0 \rightarrow \sigma_0 .
\end{equation} 
At the final stage of the calculation we rescaled solutions imposing
the condition $\Phi_{\infty}=0$ in order to fix a unique mass scale
for all of them.

 The numerical strategy consists in solving the system (12)--(16) starting
 from the series solution (22)--(26) near the origin. The crucial role
 is played by the parameter $b$ which shoul take a discrete sequence
 of values. For $\beta=0$ the solution of Eq. (32)
 reads $\Phi_2=-2b^2\exp (-2\Phi_0)$, and clearly
 this does not impose any restriction
 on this parameter. But for $\beta\neq 0$ it turns out that real
 solutions for $\Phi_2$ do not exist in some region of $b$. Hence,
 in addition to the problem of ``quantization'' of $b$ one has
 to ensure that $b$ belongs to region where the real roots of the
 Eq. (32) exist. It happens that if $\beta$ is greater than some
 ($N$--dependent) critical value $\beta_N$,
 the allowed region of $b$ does not
 contain those quantized values for which regular solutions exist.
 Only for $\beta < \beta_N$ regular
 solutions exist and exibit behaviour similar to that of the EYMD
 solutions.

 Real roots of the algebraic equation (32) form two branches as shown on the
 Fig. 3a,b in terms of the quantities
 ${\tilde b}=b\exp (-2\Phi_0),\, {\tilde \Phi_2}=\Phi_2 \exp (-2\Phi_0)$.
 For roots from the second branch (Fig. 3b) we didn't find any solution,
 they seem to correspond to $b$ outside the above quantization domain.
 Note that this branch does not contain the EYMD root corresponding to
 $\beta=0$. The first branch 3a {\em  has} a solution for $\beta=0$,
 while for any
 $\beta\neq 0$ there are two negative solutions with absolute values
 $\Phi_2^{max}(b)$ and $\Phi_2^{min}(b)$. From these two, it is
 just the second one, $\Phi_2^{min}(b)$, which has the limiting value
 $\Phi_2=-2b^2\exp (-2\Phi_0)$ when $\beta \rightarrow 0$. No regular
 solutions  to the system (12)--(16)
 corresponding to $\Phi^{max}(b)$ were found neither.

 Starting with the known $\beta=0$, $N=1$ EYMD solution \cite{dg1} we
increased gradually the value of $\beta$ searching for the desired
quantized $b$ related to $\Phi_2^{min}(b)$.
Numerical integration of the system
(12)--(16) was done using the Runge--Kutta fourth order scheme.  The
values of the parameters for $N=1$ case, found numerically for
some $\beta$ together with the corresponding  ADM mass $M=D$  are
given in the Table 1. The solutions were rescaled to ensure
$\Phi_{\infty}=0$.

 \begin{table}[htb]\begin{center}
 \centerline{\small\bf Table 1. N=1.}
 \medskip

 \begin{tabular}{|c|c|c|c|c|c|c|c|}       \hline
 $\beta$ & $b$ & $\Phi_0$ & $\sigma _0$ & $\Phi_2^{min}$ & $\Phi_2^{max}$
 & $M=D$  & $w_2 $\\  \hline
 0. & 1.073 & 0.9311 & 0.3936 & -0.3576 & --- & 0.578 &  -0.7153     \\
 0.1000& 1.026& 0.9199 & 0.3840 & -0.3523 & -3.390 & 0.573  & -0.7344  \\
 0.2000& 0.9866& 0.9122& 0.3744 & -0.3566 & -1.475 & 0.568  & -0.7697  \\
 0.3000& 0.9619& 0.9120& 0.3597 & -0.3833 & -0.8376 & 0.563  & -0.8496 \\
 0.3700& 0.9657& 0.9231& 0.3421 & -0.4938 & -0.5198 & 0.560  & -1.0933 \\
 \hline
 \end{tabular}\end{center}\end{table}
 \noindent

 One can observe that with increasing $\beta$ two real roots
 $\Phi_2^{min}(b)$ and $\Phi_2^{max}(b)$ converge and merge together for
 a limiting value  $\beta_1$ approximately equal to $0.37$.
 For $\beta>\beta_1$ there are no such $b$ which could generate
 asymptotically flat solutions with $N=1$ compatible with the existence
 of the real root $\Phi_2(b)$ of the Eq. 32.

Similar situation was encountered for higher--$N$ solutions.
 Numerical results for $N=2$ and $N=3$ are presented in the Tables 2, 3.
 Figures 4--8 depict the corresponding numerical curves for
 some values of $\beta$ and $N$.

 \begin{table}[htb]\begin{center}
 \centerline{\small\bf Table 2. N=2.}
 \medskip

 \begin{tabular}{|c|c|c|c|c|c|c|c|}       \hline
 $\beta$ & $b$ & $\Phi_0$ & $\sigma _0$ & $\Phi_2^{min}$ &
 $\Phi_2^{max}$ & $M=D$  & $w_2$ \\  \hline
 0. & 8.3612 & 1.7923 & 0.1665 & -3.8796 & ---    & 0.685  & -7.760  \\
 0.1000& 7.1902& 1.7481 & 0.1529 & -3.5165 & -15.982& 0.673  & -7.558  \\
 0.2000& 6.4017& 1.7297 & 0.1370 & -3.6597 & -5.7461& 0.660  & -8.161  \\
 0.2208& 6.3344& 1.7343& 0.1320& -4.2904 & -4.3127& 0.657 &  -9.478  \\
 \hline
 \end{tabular}\end{center}\end{table}
 \noindent

 \begin{table}[htb]\begin{center}
 \centerline{\small\bf Table 3. N=3.}
 \medskip

 \begin{tabular}{|c|c|c|c|c|c|c|c|}       \hline
 $\beta$ & $b$ & $\Phi_0$ & $\sigma _0$ & $\Phi_2^{min}$ &
 $\Phi_2^{max}$ & $M=D$  & $w_2$ \\  \hline
 0. & 53.8351 & 2.6920 & 0.0678 & -26.600 & --- & 0.7042  & -53.202     \\
 0.2000& 36.5453& 2.5930 & 0.0504 & -22.348 & -30.543  & 0.6744 & -49.817  \\
 0.21117& 36.2683& 2.5956& 0.0492 & -25.151 & -25.212 & 0.6726 & -55.558 \\
 \hline
 \end{tabular}\end{center}\end{table}
 \noindent

 Note, that the numerical values of ADM mass/dilaton charge
 monotonically decrease with growing $\beta$ for each fixed $N$.
 Also, it can be observed that the
 limiting values of $\beta_{N}$ decrease with the increasing $N$:
 ($\beta_{1}=0.37, \beta_{2}=0.22, \beta_{3}=0.21, ...$).
 It can be anticipated that $\beta_{N}$ has a limiting value
 $\beta_{\infty} $ as $N \rightarrow \infty$, which would presumably
 give an absolut bound of the existence of static spherically symmetric
 regular EYMD--Gauss--Bonnet solutions.
It is also interesting to note, that although the contribution
of the GB terms to the energy density for $\beta$ close to
$\beta_N$ is not small (as it is shown on Fig. 5),
the behaviour of $f$ and metric functions is very similar
to that of pure EYMD solutions.
 It has also to be noted that, when  the limiting value  $\beta_{N}$
 is approached, neither singularities no other
 numerical problems arise; so the only reason for the absence
 solutions  when $\beta$ exceeds the above critaical value is an intrinsic
 incompatibility of the series expansion near the origin.

We conclude with the following remarks. When Gauss--Bonnet term is included,
the total number of derivatives in the system of equations increases,
as well as the dergee of its non--linearity. However, in a
limited region of the numerical factor $\beta$ the behaviour of
solutions remains qualitatively the same as in the pure EYMD case.
Moreover, the remarkable equality of the ADM mass to the dilaton charge
remains unaffected by the GB term for any $\beta$. However it is
likely that  EYMD sphalerons are destroyed by the Gauss--Bonnet term
for sufficiently large values of $\beta$. The most persistent is the
$N=1$ solution, which exists up to $\beta=0.37$. Higher $N$ solutions
cease to exist for lower $\beta$, the limiting value is likely
to be of the order of 0.2.

  This work was supported in part by the
  ISF Grant M79000 and by the Russian Foundation
  for Fundamental Research Grant 93--02--16977.

\newpage
 {\Large \bf Figure Captions}\vskip20pt

 Fig. 1. GB term (B) and GB density (A), calculated for pure EYMD
    $N=1$ solutions \cite{dg1},
    curves (C) and (D) -- GB density for $N=2,3$ EYMD solutions.

 Fig. 2. Contributions to the energy density $r^2 * T^0_0$,
    from YMD ($\beta=0$) and GB parts ($\beta=1$)
    calculated using EYMD solutions :
    (A): $N=1$, YMD;
    (B): $N=1$, GB;
    (C): $N=2$, YMD;
    (D): $N=2$, GB.

 Fig. 3a,b. Real roots of Eq. 32 (two different branches), $\beta=0.1, 0.2,
    0.37, 0.5,1$ in terms of
    ${\tilde b}=b\exp (-2\Phi_0),\, {\tilde \Phi_2}=\Phi_2 \exp (-2\Phi_0)$.

 Fig. 4. "Gauss--Bonnet" mass distribution (contribution to ADM mass
    from $\beta$~--dependent terms) for  solutions
    with $\beta=0.2$, $N=1,2,3$.

 Fig. 5. Energy density for $N=3$, $\beta=0.2$.
    (A): total energy density; (B): contribution from
    $\beta$~-independent terms; (C): GB contribution.

 Fig. 6. Yang-Mills function $f$ for $N=1,2,3$.
    Solid lines: solutions with GB
    term ($\beta=0.2$), dashed lines: purely EYMD solutions

 Fig. 7. Metric function $W=g_{00}$ (dashed lines) and $\exp (-2\Phi)$
    (solid lines) for $N=1,2,3$,
    $\beta=0.2$.

 Fig. 8. Metric function $\sigma$ for $N=1,2,3$.
    Solid lines: solutions with GB
    term ($\beta=0.2$); dashed lines: purely EYMD solutions
     ($\beta=0$).

 \end{document}